\documentclass[sn-mathphys]{sn-jnl}% Math and Physical Sciences Reference Style
\jyear{2021}%
\usepackage{here}
\usepackage{physics}
\usepackage{soul}
\begin{document}

\title[ ]{Site-dependent control of polaritons in the Jaynes--Cummings--Hubbard model with trapped ions}


\author*[1,2]{\fnm{Silpa } \sur{Muralidharan}}\email{u928298g@ecs.osaka-u.ac.jp}

\author[2,3]{\fnm{Kenji } \sur{Toyoda}}\email{kenji.toyoda.qiqb@osaka-u.ac.jp}

\affil*[1]{\orgdiv{Graduate School of Engineering Science}, \orgname{Osaka University}, \orgaddress{\street{1-3 Machikaneyama}, \city{Toyonaka}, \postcode{5600043}, \state{Osaka}, \country{Japan}}}

\affil[2]{\orgdiv{Center for Quantum Information and Quantum Biology},\orgname{Osaka University}, \orgaddress{\street{1-2 Machikaneyama}, \city{Toyonaka}, \postcode{5600043}, \state{Osaka}, \country{Japan}}  }


\abstract{
We demonstrate the site-dependent control of polaritons in the Jaynes--Cummings--Hubbard (JCH) model with trapped ions.
In a linear ion crystal under illumination by optical beams nearly resonant to the red-sideband (RSB) transition for the radial vibrational direction,
quasiparticles called polaritonic excitations or polaritons, 
each being a superposition of one internal excitation and one vibrational quantum (phonon), 
can exist as conserved particles.
Polaritons can freely hop between ion sites 
in a homogeneous configuration,
while their motion can be externally controlled 
by modifying the parameters for the optical beams site-dependently.
We demonstrate the blockade of polariton hopping in a system of two ions
by the individual control of the frequency of the optical beams illuminating each ion.
A JCH system consisting of polaritons in a large number of ion sites
can be considered an artificial many-body system of interacting particles,
and the technique introduced here can be used to exert fine local control over such a system, 
enabling detailed studies of both its quasi-static and dynamic properties.}

\maketitle

\section{Introduction}

\label{sec:intro}

Recent advancements in quantum information science and associated 
technologies have led to the development of 
quantum information processing, including
quantum computation and quantum simulations 
\cite{feynman1982,Blatt2012,Cirac2012,Daley2022}.
In quantum simulations, physical systems that approximately simulate
naturally existing systems, 
as well as artificially designed systems, 
which may have similarities with existing systems, 
are realized in a well-controlled physical platform, and their
properties can be studied.
The Jaynes--Cummings--Hubbard (JCH) model 
\cite{Greentree2006,hartmann2006,Angelakis2007,Ivanov2009}
is simulable in various physical platforms.
The model comprises an array of 
Jaynes--Cummings (JC) systems \cite{Jaynes1963}, where each JC system
consists of an atom-like two-level system (TLS) and a harmonic oscillator
comprising a single mode of a quantized wave field.
In a system obeying the JCH model,  
a strongly interacting system of quasiparticles
called polaritons is expected to be realized,  
where each polariton is a superposition of an excitation
in the TLS and a wave quantum such as a photon or a phonon.

The JCH model was initially proposed for an array of coupled optical cavities,
such as photonic band gap cavities, each containing 
a single two-level atom \cite{Greentree2006}.
The model has been applied to other systems including 
superconducting systems \cite{Hoffman2011,Raftery2014,Fitzpatrick2017}
and trapped ions \cite{Ivanov2009,Toyo2013,Debnath2018,Li2022}.
These systems share similar properties with respect to their 
controllability,
including the site-dependent control of the model parameters. 
In the case of a JCH system of trapped ions \cite{Ivanov2009},
the ions are illuminated with optical beams nearly resonant to the red-sideband
transition for a vibrational direction, and the parameters for
the optical beams, especially their amplitudes and frequencies,
determine important system parameters such as the site-wise
JC coupling constants
and detunings of the wave fields with respect to the TLS resonances.
Angelakis {\it et al.} \cite{Angelakis2007} 
and Ivanov {\it et al.} \cite{Ivanov2009}
found that by varying the global detuning,
a quantum phase transition (or crossover) 
between a Mott-insulator and superfluid states can be observed.
This has been reproduced in experiments observing a phase crossover
\cite{Toyo2013},
where the variation of the detuning over a relatively large span
that exceeds the magnitudes of all the relevant parameters
(JC coupling constants and the phonon hopping rate) results in
a transition from one characteristic system ground state to another
with a completely different character.
In another study, site-dependent amplitude variation was utilized
to realize control of the hopping of phonons (not polaritons) 
in a system of three trapped ions \cite{Debnath2018}.
 
In this article, we propose the use of site-dependent frequency 
shifts for optical beams 
nearly resonant to the red-sideband (RSB) transition
in trapped ions,
to control the motion of polaritons in the system.
Using this technique, we realize the blockade of polariton hopping
in a system of two ion sites.
By varying the frequencies of the optical beams, the polariton hopping
to or from a particular ion site 
can be suppressed.
Hence, it is possible to
effectively decouple one ion site from the system dynamics.

The scheme introduced here is similar to
that used in the phonon blockade demonstrated by Debnath {\it et al.}
\cite{Debnath2018},
in that both schemes utilize the discrepancy of the relevant eigenenergies
across ion sites to realize the blockade of the motion of quasiparticles
(either polaritons or phonons)
in a linear ion crystal.
In comparison to the scheme in \cite{Debnath2018},
we find that in the scheme presented here, the leakage of quasiparticles
to the ion site illuminated by the blockade optical beam
can be lower by a factor of approximately four in an optimum case, 
assuming the same RSB Rabi frequency.

The technique presented here can be applied to various studies,
and we consider here that of quantum phase crossovers
and their dependence on the number of ion sites.
There is a certain experimental cost in changing the number of ion sites,
since doing this
typically requires a reloading (or at least an additive loading when
increasing the number).
Even a change of the optical beam positions may be required
if each ion is addressed with a different beam.
By using the technique introduced here,
it is possible to decouple some of the ion sites,
and hence, effectively reduce the number of ion sites
without changing the configuration of the ion crystal itself.
This can be done by selectively shifting the optical beam frequencies
for the ions placed close to both edges of the linear crystal,
where it is assumed that
each ion is separately addressed with one dedicated optical beam.
In this way, the number dependence of phase crossovers,
in analogy to a spin system \cite{Islam2011},
can be 
observed without reloading or additive loading.
The technique introduced here can also be applied to studies
of dynamical properties
\cite{Cirac2012,Daley2022,Wong2011,Nissen2012,Li2021}, 
where strong site-dependent as well as time-dependent
disturbances can be applied to the system.

\section{Principles of polaritons in the JCH model}
\label{sec:jchmodel}

\begin{figure}
    \centering
    \includegraphics[width=7.5cm]{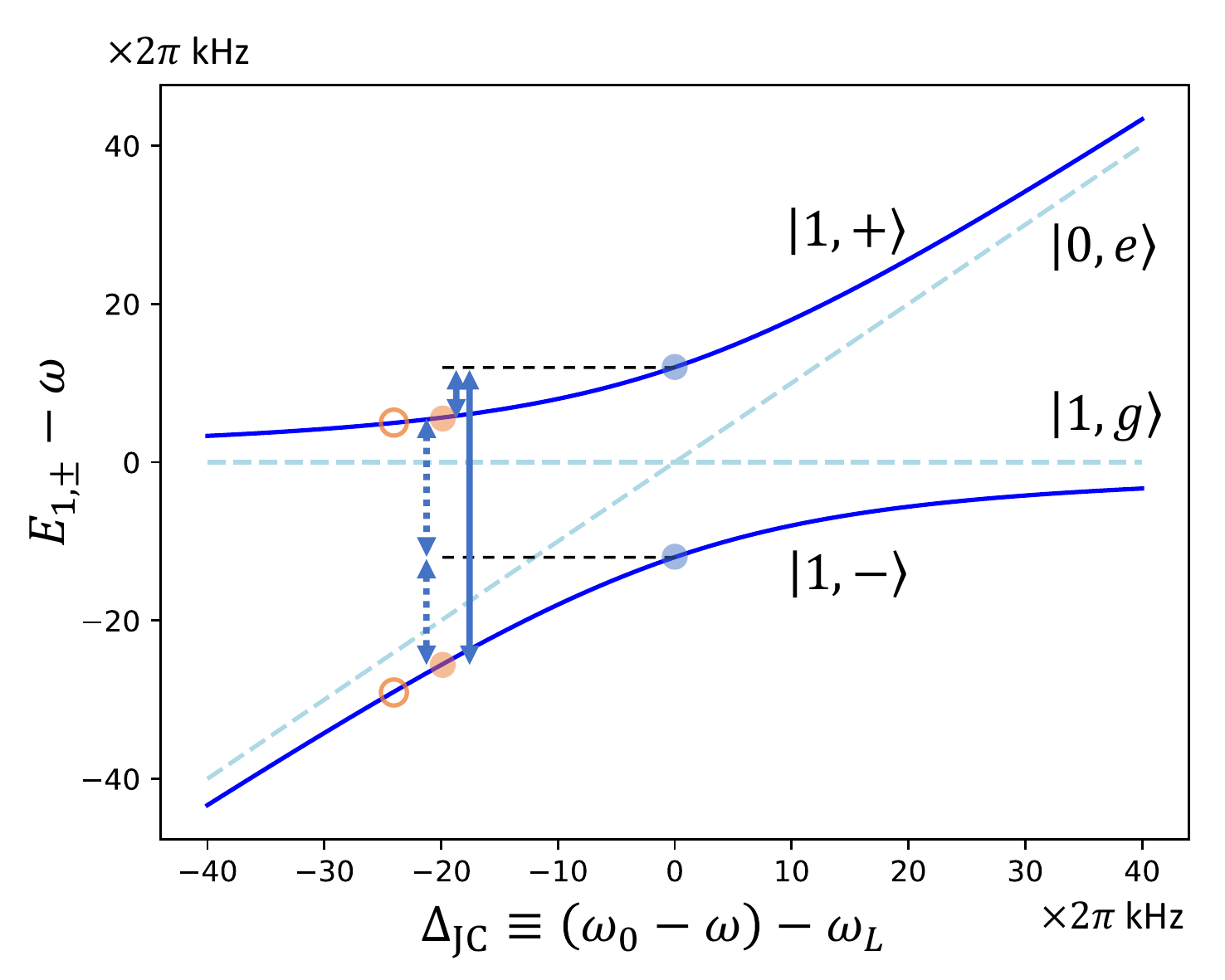}
    \caption{
Eigenenergies for dressed states in the JC model.
The eigenenergies of the dressed states $\ket{1,\pm}$ are
plotted against the JC detuning $\Delta_\mathrm{JC}$ as
blue solid curves.
The energies for the bare states $\ket{1,g}, \ket{0,e}$ are
plotted as light blue dashed lines.
For those eigenenergies,
the reference point is taken at $\omega$, 
which corresponds to
the energy of
the bare states $\ket{1,g}, \ket{0,e}$ at resonance.
The blue filled circles represent
eigenenergies at resonance, which correspond to
those at Ion 1.
The red filled circles represent
eigenenergies for $\Delta_\mathrm{JC}=-2\pi\times20$ kHz,
which correspond to those at Ion 2 in the case shown in
Fig.~\ref{fig:simex}(e), (f) and
Fig.~\ref{fig:simleakage}(a).
The blue solid arrows (blue dotted arrows) represent the energy gaps between
the $\ket{1,+}$ ($\ket{1,-}$) eigenenergy at Ion 1
and eigenenergies at Ion 2. 
The red hollow circles represent
eigenenergies for $\Delta_\mathrm{JC}=-2\pi\times24$ kHz,
which corresponds to those at Ion 2 in the case
of minimum leakage, as is shown in
Fig.~\ref{fig:simleakage}(b)}
    \label{fig:eigenenergies}
\end{figure}

In this section, 
we review the formalism for polaritons in the JCH model.
First, 
the JC Hamiltonian is introduced to describe a single ion site comprising
two internal states and a quantized radial vibrational mode
(we assume $\hbar=1$ throughout the paper):
\begin{eqnarray}
 H_\mathrm{JC} &=& 
  \omega \hat{a}^{\dagger}\hat{a} + 
  (\omega + \Delta_\mathrm{JC})
  \hat{\sigma}^{+}\hat{\sigma}^{-} \nonumber\\
 &&+g_\mathrm{JC}(\hat{a}\hat{\sigma}^{+} + \hat{a}^\dagger\hat{\sigma}^{-}).
  \label{eqn:JC}
\end{eqnarray}
Here, 
we assume that 
a laser with a frequency of $\omega_\mathrm{L}$
that is nearly resonant
to the RSB transition 
for the radial vibrational motion
is applied,
and the Hamiltonian is taken in a frame rotating with
that frequency.
$\omega$ is the vibrational frequency.
$\hat{a}^\dagger$ and  $\hat{a}$ are the creation and annihilation operators 
for radial phonons, respectively,
where the radial phonon states are assumed to be
spanned by phonon number states $\ket{n}$ ($n=0,1,\ldots$).
$\Delta_\mathrm{JC} \equiv (\omega_0 - \omega) - \omega_\mathrm{L}$
is the detuning for the JC coupling, 
where 
$\omega_0$ is the resonance frequency for the internal states.
$\hat{\sigma}^{+}$ and $\hat{\sigma}^{-}$ are the raising and lowering 
operators for the internal states,
respectively,
where the internal states are assumed to be 
spanned by the ground ($\ket{g}$) and
excited ($\ket{e}$) states.
$g_\mathrm{JC}$ is the JC coupling strength,
which is dependent on the RSB Rabi frequency. 
The first term in Eq.~(\ref{eqn:JC}) corresponds to the phonon energy
and
the second term to the atomic internal energy.
The last term represents the JC coupling induced by the RSB excitation laser.
In this system, the concept of dressed atoms \cite{CohenTannoudji2005} 
emerges, 
which has a close relation to the polaritons dealt with in this work. 
The ground state of the combined system ($\ket{0,g}\equiv\ket{0}\otimes\ket{g}$) 
and the dressed states ($\ket{p,\pm}$) possess the following eigenenergies
($p=1,2,\ldots$):
\begin{eqnarray}
 E_{0}&=&0,\\
 E_{p,\pm}&=&\omega{p}\pm\Omega_{p}(\Delta),
\label{eqn:eigenenergies2}
\end{eqnarray}
where $\Omega_{p}(\Delta) = [ \Delta ^{2} + 4g_\mathrm{JC}^{2}p)]^{1/2}$ 
is the generalized Rabi frequency.
%The energy splitting for the two states with the same quantum number 
%$p$
% is proportional to the square root of $p$,
%and thus the dependence is nonlinear. 
The eigenenergies of $\ket{1,\pm}$ are plotted against the
JC detuning $\Delta_\mathrm{JC}$ in Fig.~\ref{fig:eigenenergies}.

The combined system of the internal states of ions in a linear crystal
and their vibrational states along a radial direction under the illumination
of optical beams nearly resonant to the RSB transition
can be described by the JCH model
\cite{Ivanov2009}.
Polaritons in this system can be defined in a way similar to
dressed atoms in the single-site case,
but with the additional property of being able to propagate within
the crystal \cite{Ohira2021b}, 
which is inherited from the similar property of 
radial phonons in a linear ion crystal 
\cite{Porras2004,Haze2012,Tamura2020}.
The JCH Hamiltonian for a linear crystal with $N$ ions is given as
\begin{eqnarray}
 H_\mathrm{JCH} &=& 
  \sum_{i=1}^{N} 
  \left[
  \omega_{i}\hat{a}_{i}^{\dagger}\hat{a}_{i} + 
  (\omega_i + \Delta_{\mathrm{JC},i})
  \hat{\sigma}_{i}^{+}\hat{\sigma}_{i}^{-}
  \right]
  \nonumber\\
  &&+
  \sum_{i=1}^{N} 
  {g_i} 
  (\hat{a}_{i}\hat{\sigma}_{i}^{+}+\hat{a}_{i}^{\dagger}\hat{\sigma}_{i}^{-})
  \nonumber\\
  &&+ 
  \sum _{i< j}^{N}
  \frac{k_{ij}}{2}
  (\hat{a}_{i}\hat{a}_{j}^{\dagger} + \hat{a}_{i}^{\dagger}\hat{a}_{j}).
\label{eqn:JCH1}
\end{eqnarray}
Here, 
$\omega_i = \omega- \sum_{i\neq j}^{N} k_{ij}/2$ ($i=1,2,\ldots,N$) 
is the site-dependent vibrational frequency 
incorporating inter-ion Coulomb couplings \cite{Porras2004,Zhu2006}. 
$\hat{a}_{i}^{\dagger}$ and  $\hat{a}_{i}$ are 
the creation and annihilation operators 
for local phonons along the radial direction of the $i$th ion. 
$\Delta_{\mathrm{JC},i} \equiv (\omega_0 - \omega_i) - \omega_{\mathrm{L},i}$
is the site-dependent detuning for the JC coupling,
where
$\omega_{\mathrm{L},i}$ is the site-dependent frequency of the optical beam
illuminating the $i$th ion.
$\hat{\sigma}_{i}^{+}$ and $\hat{\sigma}_{i}^{-}$ are
the raising and lowering operators for the internal states of the $i$th ion,
respectively.
$g_i$ is the JC coupling strength for each ion site. 
The last term in Eq.~(\ref{eqn:JCH1})
describes the hopping of phonons between different ion sites.
$k_{ij}= e^2/4\pi \epsilon_0 m d_{ij}^{3}\omega_i$ 
is the hopping rate between the $i$th and $j$th ion sites
\cite{Porras2004,Haze2012},
where $m$ is the mass of an ion 
and $d_{ij}$ is the distance between the $i$th and $j$th ions.

In the case of two ions, which is relevant in the present work, 
the JCH Hamiltonian is as follows:
\begin{eqnarray}
 H_\mathrm{JCH} &=& 
  \sum_{i=1}^{2}
  \left[
   \omega\hat{a}_{i}^{\dagger}\hat{a}_{i} + 
   (\omega + \Delta_{\mathrm{JC},i})
   \hat{\sigma}_{i}^{+}\hat{\sigma}_{i}^{-}
  \right] 
  \nonumber\\
 &&+
  \sum_{i=1}^{2}
   {g_i}
   (\hat{a}_{i}\hat{\sigma}_{i}^{+}+\hat{a}_{i}^{\dagger}\hat{\sigma}_{i}^{-})
  \nonumber\\
 &&+
  \frac{k_{12}}{2}
  (\hat{a}_{1}\hat{a}_{2}^{\dagger} + \hat{a}_{1}^{\dagger}\hat{a}_{2}).
\label{eqn:JCH2}
\end{eqnarray}
Here, 
the site-dependent corrections for the vibrational frequencies
$- \sum_{i\neq j}^{N} k_{ij}/2$
are omitted and each $\omega_i$ is represented by the same symbol $\omega$
considering the symmetry in the system.

\section{Experimental procedure}
\label{sec:misf}

In the present experiment, two $^{40}$Ca$^{+}$ ions are used, 
which are trapped inside a linear Paul trap 
with an ac electric field oscillating at 24 MHz and a dc electric field. 
The single-ion vibrational frequencies for the trap 
are $(\omega_x, \omega_y, \omega_z)/2\pi= (2.95, 2.74, 0.18 )$ MHz. 
The distance between the two ions is 18 $\mu$m 
and the phonon hopping rate is $k_{12}=2\pi\times5.9$ kHz. 

Two lasers at 423 and 370 nm are used for ion loading with optical ionization.
Laser beams at 397 and 866 nm, nearly resonant to 
$S_{1/2}$--$P_{1/2}$ and 
$D_{3/2}$--$P_{1/2}$, respectively,
 are used for Doppler cooling. 
A laser at 729 nm nearly resonant to
$D_{5/2}$--$P_{3/2}$
is used for sideband cooling (SBC) and the initialization of a polariton,
as well as in the experimental step for the observation of polariton dynamics
(we call this the {\it quantum simulation step}).
The output of the 729-nm laser is divided and
two beams are applied to the two ions so that
each ion is illuminated with one dedicated beam.
The resulting maximum carrier (RSB) Rabi frequency is 
measured to be $\sim600$ (24) kHz for each ion. 
A laser at 854 nm is used to deplete the population in $D_{5/2}$
during SBC and after the manipulation of polaritons.
The detection of the internal states
is done by illumination with 397- and 866-nm pulses and by collecting
fluorescence photons with a photomultiplier tube.

In our setup, the frequencies for the optical pulses at 729 nm illuminating 
the respective ions (Ions 1 and 2) 
during the quantum simulation step
can be set to be either equal to or different from each other,
and in the latter case the finite frequency difference
may lead to a blockade of polariton hopping.
In the present experiment, the frequency for the pulse illuminating
Ion 1 is kept at the resonance of the RSB transition,
while that for the pulse illuminating Ion 2 is varied over a certain range.

The frequencies for the optical pulses
are controlled using
acousto-optic modulators (AOMs) in both 
double-pass and single-pass configurations.
A double-pass AOM system \cite{Donley2005} is used for the base frequency
control for all the optical beams at 729 nm,
while a single-pass AOM system is used to control
the frequency of the respective optical beam illuminating each ion.
In contrast to a double-pass configuration,
a single-pass configuration enables amplitude/frequency modulation
with high efficiency in the diffracted power, though
an angle shift of the diffracted beam and a resulting
spatial shift of the optical beam at the position of the ions
cannot be avoided.
In the present experiment with a maximum frequency shift of $\sim$20 kHz,
the spatial shift is estimated to be $\sim$0.1 $\mu$m,
which is sufficiently small compared with the beam waist size 
($e^{-2}$ radius $\sim$3 $\mu$m).

\begin{figure}
    \centering
    \includegraphics[width=7.5cm]{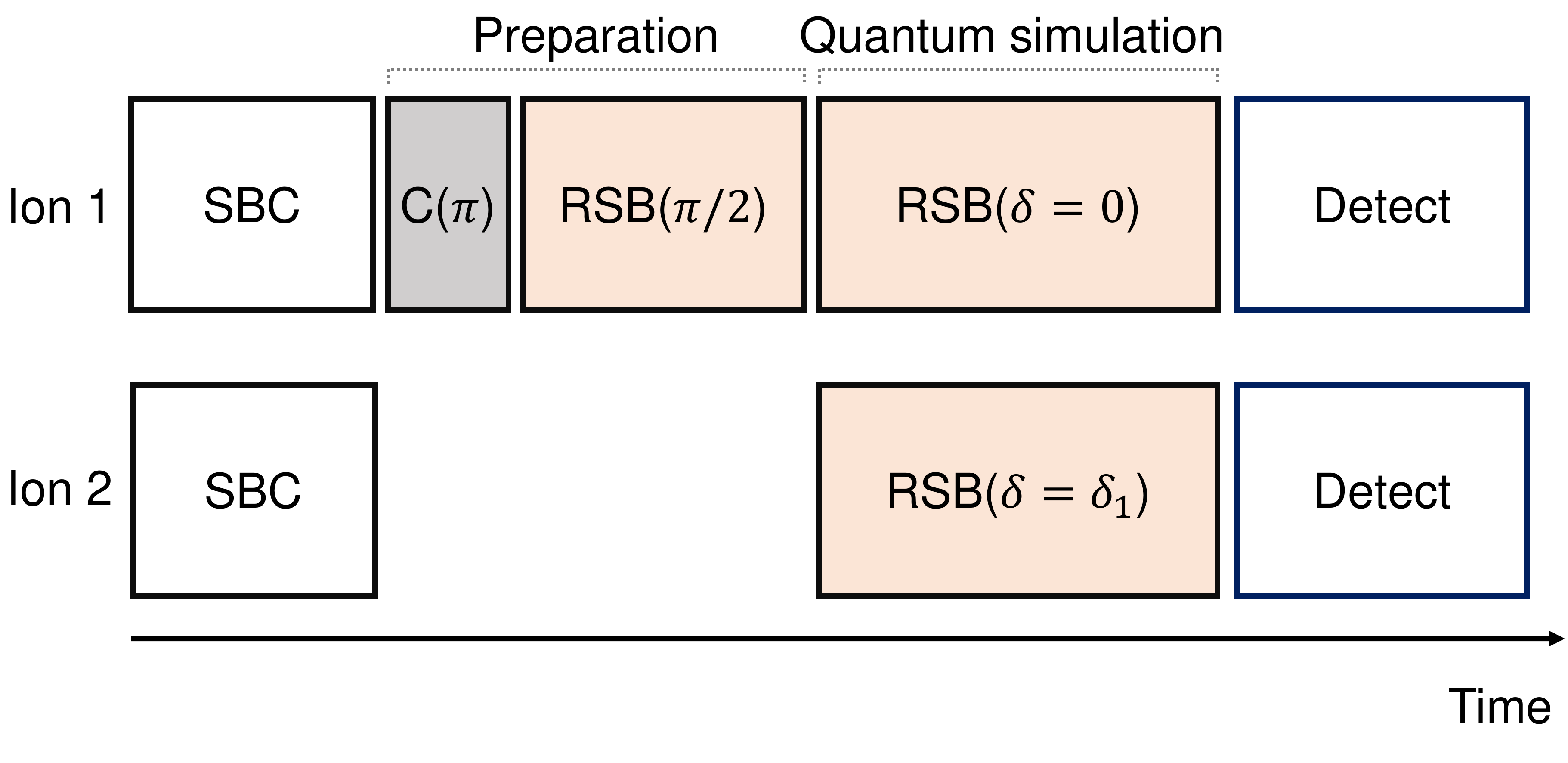}
\caption{ Time sequence for observing
    polariton blockade by individual control of the optical beam frequency.
$\mathrm{C}(\pi)$ represents a carrier $\pi$ pulse.
$\mathrm{RSB}(\pi/2)$ represents a resonant RSB pulse with 
pulse area $\pi/2$.
$\mathrm{RSB}(\delta)$ represents either a resonant ($\delta=0$) 
or off-resonance ($\delta\neq0$) RSB pulse with a certain length
}
\label{fig:pulse}
\end{figure}

Figure~\ref{fig:pulse}
shows the time sequence for the present experiment (after Doppler cooling).
As is described above, we use lasers at 397 and 866 nm to perform
Doppler cooling. 
Further cooling is obtained by 
performing SBC with lasers at 729 and 854 nm, 
thereby cooling the ions to near the vibrational ground state
in the radial directions. 
The average phonon numbers in the radial directions measured after SBC are
$(\langle{n}_x\rangle,\langle{n}_y\rangle)=(0.30,0.03)$. 
Then, the initial state for the observation of polariton hopping and
a blockade is prepared.
This is done by applying 
a carrier $\pi$ pulse (length $\sim{1}$ $\mu$s)
and RSB $\pi/2$ pulse (length $\sim{10}$ $\mu$s) in succession
to one of the two ions, which is identified as Ion 1 (and the other as Ion 2).
The state in Ion 1 after these is a superposition of 
$\ket{0,e}\equiv\ket{0}\otimes\ket{e}$ and 
$\ket{1,g}\equiv\ket{1}\otimes\ket{g}$,
where 
$\ket{g}$ corresponds to 
the internal ground  state ($\ket{S_{1/2},m_J =-1/2}$)
and
$\ket{e}$ to the internal excited state ($\ket{D_{5/2},m_{J'}=-1/2}$).
This state corresponds to one polariton in Ion 1.
The preparation step just described is followed by 
the quantum simulation step, where the ions are
illuminated with RSB optical pulses that have
equal intensities with respect to both ions,
while their frequencies are adjusted depending on
which type of experiment we perform, {\it i.e.},
without a frequency difference for no blockade
and with a finite frequency difference in the case of observing 
a polariton blockade.
The time sequence is repeated while varying the length for
the quantum simulation step from zero to a certain length
so that the time-dependent behavior of the system is acquired.

\section{Results}
\label{sec:condmeas}

\begin{figure*}[h]
    \centering
    \includegraphics[width=\textwidth]{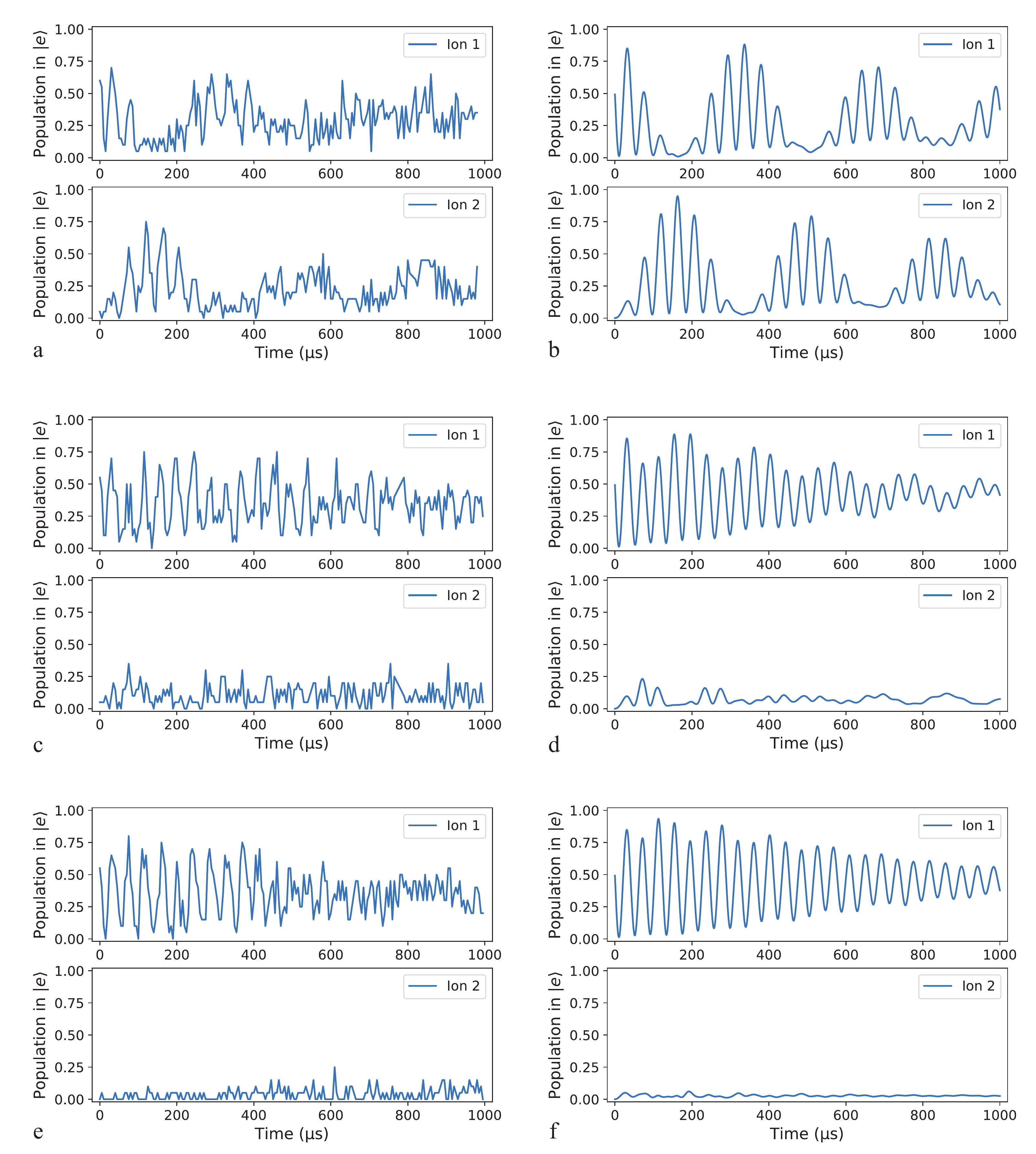}
    \caption{Observation of polariton hopping and its blockade by
illuminating individually addressed RSB optical pulses with
site-dependent frequency shifts.
(a) Experimental result for zero detuning ($\Delta_{\mathrm{JC},2}=0$) and
(b) corresponding numerically calculated result, where
a relatively fast oscillation
with a period of $\sim$40 $\mu$s
and a relatively slow change of the envelope
with the period of $\sim$350 $\mu$s
can be seen.
(c) Experimental result for $\Delta_{\mathrm{JC},2}=-2\pi\times10$ kHz and
(d) corresponding numerical result, where
a relatively fast oscillation still exists while
the relatively slow oscillation of the envelope is not apparent.
(e) Experimental result for $\Delta_{\mathrm{JC},2}=-2\pi\times20$ kHz and
(f) corresponding numerical result,
where again the relatively slow oscillation of the envelope is not apparent
and the internal excited state population at Ion 2 is largely suppressed
}
    \label{fig:simex}
\end{figure*}

The results for the observation of polariton hopping
and the blockade of a polariton using frequency
shifts of individually addressing RSB pulses
are shown in Fig.~\ref{fig:simex}.
In Fig.~\ref{fig:simex}(a),
the population in $\ket{e}$ in the case without
a frequency difference in the RSB pulses between the two ion sites
is plotted against elapsed time.
After a polariton is prepared in Ion 1 (before the time $t=0$),
it starts hopping to the other ion site (Ion 2)
to form a superposition between the two.
Still, part of its wave packet remains
in the initial ion site for a while ($\sim100$ $\mu$s),
manifesting as
a relatively fast oscillation with a period of $\sim$40 $\mu$s.
This oscillation is observed because 
the polariton state does not coincide with
one of the eigenstates of the instantaneous JC Hamiltonian
under the illumination of the RSB pulses,
but in fact corresponds to a superposition of eigenstates,
that is, $\ket{1,+}$ and $\ket{1,-}$,
and the interference between these two states generates
the relatively fast oscillation in the population 
observed before $t=100$ $\mu$s.
We can also describe the fast oscillation as a sideband Rabi oscillation
induced by the RSB pulse.
In any case, the structure can be understood as local time-dependent
behavior within one ion site.
Eventually, the polariton moves to Ion 2, which
is confirmed by the oscillatory behavior
observed for Ion 2 in $t=0$--$300$ $\mu$s.
From around $t$ = 200 $\mu$s, a similar oscillatory behavior is again 
seen in Ion 1 (along with larger noise), 
which corresponds to the wave packet
returning to the original ion site.
In this way, the polariton 
goes back and forth between the two ion sites, and this behavior
is confirmed in the envelopes of the populations in Ions 1 and 2,
which again show oscillatory behaviors 
with a longer period of $\sim$340 $\mu$s.
The oscillations in the envelopes show relaxations
due to decoherence in the system.
The decoherence can be ascribed to such factors as 
intensity fluctuations of the optical beam
and associated frequency fluctuations due to AC Stark shifts,
as well the finite temperature after performing SBC.
The causes for the decoherence in our system are discussed in more detail
in \cite{Ohira2022}.

A numerically calculated result corresponding to the experimental result
in Fig.~\ref{fig:simex}(a)
is shown in Fig.~\ref{fig:simex}(b). 
In the numerical calculation, 
we assume 
$2g_1=2g_2=2\pi\times24$ kHz and $k_{12}=2\pi\times5.9$ kHz.
We also incorporate the error due to the decoherence \cite{Ohira2022}.
For this, we assume
a thermal distribution in the radial $y$ direction
with $\langle{n}_y\rangle=0.03$, and 
relative intensity fluctuations with a standard deviation
of $\sim0.02$.
It is confirmed from the results that
the qualitative behavior of the experimental result is 
well reproduced in the calculated result.

Next, we investigate the effect of changing the detuning of the RSB pulse for Ion 2.
The frequency of the RSB pulse illuminating Ion 2 is
increased by 10 kHz 
($\Delta_{\mathrm{JC},2}=-2\pi\times10$ kHz),
whereas that illuminating Ion 1 is kept to be resonant
($\Delta_{\mathrm{JC},2}=0$).
The result is shown in Fig.~\ref{fig:simex}(c).
Now, both the slow oscillation of the envelope and the fast oscillation
within the convex structures 
in the population for Ion 2
are much less discernible compared with
the results explained above. 
On the other hand, the fast oscillation in the population
for Ion 1 is clearly seen, and its envelopes now hardly oscillate,
although the amplitude of the fast oscillation decays
due to decoherence.  
A similar trend is confirmed in the case of
a frequency difference of 20 kHz
($\Delta_{\mathrm{JC},2}=-2\pi\times20$ kHz)
in Fig.~\ref{fig:simex}(e),
where the population in Ion 2 is further suppressed.
The qualitative behaviors in the case of finite frequency differences
are again reproduced in
the corresponding numerical results (Fig.~\ref{fig:simex}(d) and (f)).

The results in Fig.~\ref{fig:simex}
indicate that, by detuning the RSB pulse applied to Ion 2,
it is possible to make a substantial part of the polariton wave packet 
remain in the original ion site.
This is confirmed by, e.g., the sustaining amplitude of the fast oscillation
at Ion 1 observed in Fig.~\ref{fig:simex}(c) and (e).

We should note that
the stray population in $\ket{e}$ at Ion 2 shown in the figure
(i.e., the small oscillating or flat background population seen 
in the lower rows of Fig.~\ref{fig:simex}(c), (d), (e) and (f))
does not directly correspond to that of polaritons.
To evaluate the polariton population correctly, 
a simultaneous observation of 
both the internal and motional states is required,
which necessitates a specially designed observation scheme,
as is investigated in \cite{Muralidharan2021}.
We do not follow such a scheme in the present work,
and hence, the information about the polaritons
in Ion 2 is limited here.

To support the argument above,
we perform numerical calculations to 
estimate the polariton population in the next section,
where we discuss the quality of polariton blockade by estimating
the leakage of the polariton population to Ion 2.  

\section{Discussion}
\label{sec:discussion}

In this section, we discuss quantitatively the performance of 
polariton blockade by utilizing numerical calculations.

\subsection{Estimation of leakage}

\begin{figure}
    \centering
    \includegraphics[width=7.5cm]{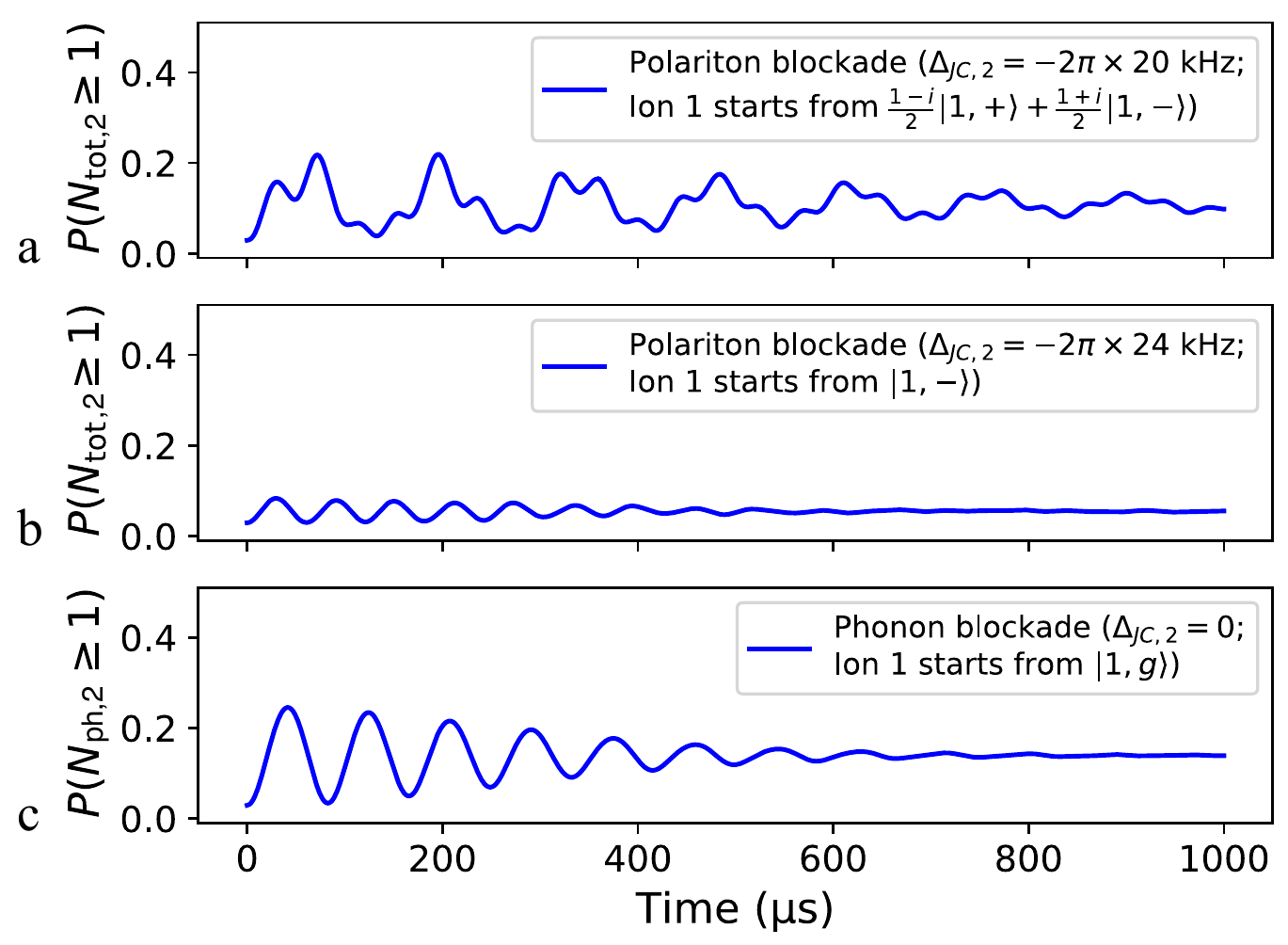}
    \caption{Numerically calculated results for ({\bf a}),({\bf b}) polariton
and ({\bf c}) phonon populations in the respective excited states at Ion 2,
plotted against elapsed time.
 In ({\bf a}) and ({\bf b}), the total population in the 
states with polariton number of one or greater,
$P({N}_{\mathrm{tot},2}\geq1)$, for 
$\Delta_{\mathrm{JC},2}=-2\pi\times20$ kHz
and $-2\pi\times24$ kHz, respectively
and the initial state at Ion 1
of $(\ket{0,e}-i\ket{1,g})/\sqrt{2}=\frac{1-i}{2}\ket{1,{+}}+\frac{1+i}{2}\ket{1,{-}}$
and $\ket{1,-}$
is plotted.
 In ({\bf c}), the total population in the 
states with the phonon number of one or greater,
$P({N}_{\mathrm{ph},2}\geq1)$,
is plotted
}
    \label{fig:simleakage}
\end{figure}

Figure~\ref{fig:simleakage}(a) shows
the calculated probability that one or more polaritons exist in Ion 2 (${N}_{\mathrm{tot},2}\geq1$), where
the conditions for the numerical calculation is set to be
identical to that in Fig.~\ref{fig:simex}(f).
Here, 
${N}_{\mathrm{tot},2}$ is the quantum number for
$\hat{N}_{\mathrm{tot},2}\equiv\hat{a}_{2}^{\dagger}\hat{a}_{2}+\hat{\sigma}_{2}^{+}\hat{\sigma}_{2}^{-}$,
which represents the polariton number at Ion 2.
We ignore the case where the polariton number per ion site is
greater than or equal to 3,
and thus, the sum of the populations for which either 
${N}_{\mathrm{tot},2}$ is 1 or 2 is displayed in 
Fig.~\ref{fig:simleakage}(a).
The initial state prepared in Ion 1 is described as
$(\ket{0,e}-i\ket{1,g})/\sqrt{2}=\frac{1-i}{2}\ket{1,{+}}+\frac{1+i}{2}\ket{1,{-}}$.
The average population over the displayed time range,
which we consider to be one measure for the degree of leakage,
is obtained to be 0.109 in this case. 
The instantaneous population even exceeds 0.2 at some points.

We have found that
the amount of leakage as given in the last paragraph
can be suppressed by choosing the appropriate conditions,
that is, those for the detuning at Ion 2 and the initial state at Ion 1.
Figure~\ref{fig:simleakage}(b) displays a result for such a case.
Here, the detuning $\Delta_\mathrm{JC,2}$ is set to be $-2\pi\times24$ kHz,
and   
$\ket{1,{-}}$ is chosen as the initial state.
We find lower leakage for the Ion 1 initial state of $\ket{1,{-}}$, 
compared with $\ket{1,{+}}$
(or any superpositions of those), around this value of the detuning.
The amount of leakage at this point is obtained to be 0.0549.
At this detuning,
the eigenenergy of 
$\ket{1,{-}}$ at Ion 1, which is calculated to be $-2\pi\times11.8$ kHz,
is different from the eigenenergies for Ion 2
($-2\pi\times28.8$ kHz and $2\pi\times4.8$ kHz, represented
by red circles in Fig.~\ref{fig:eigenenergies});
that is, there exist finite energy gaps
between the eigenenergies at Ion 1 and Ion 2. 
This condition is not satisfied to the same extent for detunings
in other regions,
%different from this,
and either of the two eigenenergies at Ion 2 become closer to that
of $\ket{1,{-}}$ at Ion 1
(see also Table~\ref{table:Delta_minus} for the relevant eigenenergies
as well as the amount of leakage against various negative values for the detuning).
If the detuning is decreased from $-2\pi\times24$ kHz,
we observe that the leakage increases.
We speculate that the energy gaps, as introduced above, can approximately 
explain the behavior that the leakage takes the minimum value
at around $-2\pi\times24$ kHz.
We can confirm from Table~\ref{table:Delta_minus}
that the smaller of the two relevant energy gaps,
$E_{1,-,1}-E_{1,-,2}$ and $E_{1,+,2}-E_{1,-,1}$
($E_{1,\pm,i}$ represents the eigenenergy of $\ket{1,\pm}$ at the
$i$th ion),
takes the largest value ($2\pi\times16.6$ kHz) 
at $\Delta_\mathrm{JC,2}=-2\pi\times24$ kHz.
We should note that the eigenenergies shown in Fig.~\ref{fig:eigenenergies}
and Table~\ref{table:Delta_minus}
incorporate only local JC interactions and not the hopping term,
and hence, the above-mentioned explanation 
holds only approximately.

\begin{table*}
    \centering
\caption{Eigenenergies, their differences and the 
amount of leakage against the negative values of the
detuning $\Delta_{\mathrm{JC},2}$. 
The values for the detuning and the eigenenergies have units of $2\pi\times$kHz.
It is assumed that $E_{1,-,1}\sim-2\pi\times11.8$ kHz}
\begin{tabular}{rrrrrrrrrr}
 \toprule
 $\Delta_{\mathrm{JC},2}$ & $E_{1,-,2}$ & $E_{1,+,2}$ & $E_{1,-,1}-E_{1,-,2}$ & $E_{1,+,2}-E_{1,-,1}$ & $\langle{}P({N}_{\mathrm{tot},2}\geq1)\rangle$ \\
 \midrule
$-$500 & $-$500.3 & 0.3 & 488.5 & 12.0 & 0.0805 \\
$-$100 & $-$101.4 & 1.4 & 89.6 & 13.1 & 0.0718 \\
$-$50 & $-$52.6 & 2.6 & 40.9 & 14.4 & 0.0631 \\
$-$30 & $-$34.1 & 4.1 & 22.3 & 15.8 & 0.0563 \\
$-$28 & $-$32.3 & 4.3 & 20.5 & 16.0 & 0.0557 \\
$-$26 & $-$30.5 & 4.5 & 18.8 & 16.3 & 0.0551 \\
$-$24 & $-$28.8 & 4.8 & 17.0 & 16.6 & 0.0549 \\
$-$22 & $-$27.1 & 5.1 & 15.3 & 16.9 & 0.0551 \\
$-$20 & $-$25.4 & 5.4 & 13.7 & 17.2 & 0.0560 \\
$-$15 & $-$21.4 & 6.4 & 9.7 & 18.2 & 0.0628 \\
$-$10 & $-$17.8 & 7.8 & 6.0 & 19.5 & 0.1040 \\
 \bottomrule
\end{tabular}
\label{table:Delta_minus}
\end{table*}

\begin{table*}
    \centering
\caption{Eigenenergies, their differences and the 
amount of leakage against the positive values of the
detuning $\Delta_{\mathrm{JC},2}$.
The values for the detuning and the eigenenergies have units of $2\pi\times$kHz
It is assumed that $E_{1,+,1}\sim2\pi\times11.8$ kHz}
\begin{tabular}{rrrrrrrrrr}
 \toprule
 $\Delta_{\mathrm{JC},2}$ & $E_{1,-,2}$ & $E_{1,+,2}$ & $E_{1,+,1}-E_{1,-,2}$ & $E_{1,+,2}-E_{1,+,1}$ & $\langle{}P({N}_{\mathrm{tot},2}\geq1)\rangle$ \\
 \midrule
10 & $-$7.8 & 17.8 & 19.5 & 6.0 & 0.1038 \\
15 & $-$6.4 & 21.4 & 18.2 & 9.7 & 0.0627 \\
20 & $-$5.4 & 25.4 & 17.2 & 13.7 & 0.0558 \\
22 & $-$5.1 & 27.1 & 16.9 & 15.3 & 0.0550 \\
24 & $-$4.8 & 28.8 & 16.6 & 17.0 & 0.0545 \\
26 & $-$4.5 & 30.5 & 16.3 & 18.8 & 0.0547 \\
28 & $-$4.3 & 32.3 & 16.0 & 20.5 & 0.0553 \\
30 & $-$4.1 & 34.1 & 15.8 & 22.3 & 0.0560 \\
50 & $-$2.6 & 52.6 & 14.4 & 40.9 & 0.0626 \\
100 & $-$1.4 & 101.4 & 13.1 & 89.6 & 0.0712 \\
500 & $-$0.3 & 500.3 & 12.0 & 488.5 & 0.0799 \\
 \bottomrule
\end{tabular}
\label{table:Delta_plus}
\end{table*}

The amount of leakage for the positive values of
the detuning $\Delta_\mathrm{JC,2}$
along with the relevant eigenenergies
are summarized in Table~\ref{table:Delta_plus}.
For positive detuning, the initial state of $\ket{1,+}$ at Ion 1 gives
a lower leakage to Ion 1, and hence,
only the results in this case are shown in the table
and the case of $\ket{1,-}$ is omitted.
We confirm that the dependence of the leakage against the detuning
shows an approximately symmetric behavior when compared with
the case of negative detuning (Table~\ref{table:Delta_minus}).
Qualitatively, this can be well understood from the symmetry of
the eigenenergy values, as is seen in
Fig.~\ref{fig:eigenenergies}, Table~\ref{table:Delta_minus}
and Table~\ref{table:Delta_plus}.

\subsection{Comparison with phonon blockade}
We also estimate the leakage in the case of phonon hopping,
as studied in \cite{Debnath2018}, for comparison
with the polariton case studied here.
Formally, the phonon hopping case corresponds to
the assumption that $\Delta_\mathrm{JC,2}=0$ and $\ket{1,g}$ is used for
the initial state at Ion 1.
We use the same value of the JC coupling constant at Ion 2 as before, 
$2g_2=2\pi\times24$ kHz.
The numerically calculated result for the time dependence of 
$P({N}_{\mathrm{ph},2}\geq1)$ in this case
is shown in Fig.~\ref{fig:simleakage}(c),
where ${N}_{\mathrm{ph},2}$ is the quantum number for
$\hat{N}_{\mathrm{ph},2}\equiv\hat{a}_{2}^{\dagger}\hat{a}_{2}$
representing the phonon number at Ion 2.
The average of $P({N}_{\mathrm{ph},2}\geq1)$ over the displayed
time range is obtained to be 0.139.

So far, we have assumed that
the phonon system has a finite thermal distribution
(we have assumed $\langle{n}_y\rangle=0.03$ in the numerical calculations).
To precisely compare the cases for polaritons and phonons,
we should use the amounts of leakage
in the cases without the thermal distribution. 
We have also performed numerical calculations for this case, 
and the results are as follows:
for a polariton blockade with 
$\Delta_\mathrm{JC,2}=-2\pi\times24$ kHz
and $\ket{1,-}$ as the initial state at Ion 1,
the amount of leakage is 0.0297,
while for a phonon blockade it is 0.1144.
Therefore, it is expected that, for a polariton blockade
based on the scheme presented in this work,
the amount of leakage is smaller than
the case of phonon hopping by approximately four times, assuming the same intensity of the optical
beam at Ion 2.

\section{Conclusions}
\label{sec:conclusions}

In conclusion,
we have demonstrated
the site-dependent control of polaritons in the Jaynes--Cummings--Hubbard (JCH) model with trapped ions.
We have presented experimental results
on blocking the hopping of a polariton initialized
in an ion site to an adjacent site,
by shifting the frequency of the optical beam illuminating the latter.
We have also performed numerical calculations to 
estimate the leakage to the adjacent ion site.
Based on the calculated results, we have found that
the gaps between the relevant eigenenergies at both ion sites
may partly explain the dependence of the amount of leakage
on the frequency shift.
The scheme presented here
can be applied to various studies on polariton systems in the JCH model,
including quasi-static as well as nonequilibrium behaviors.

\backmatter

\bmhead{Acknowledgments}

This article was supported by MEXT Quantum Leap Flagship Program (MEXT Q-LEAP) Grant Number JPMXS0118067477.

\section*{Declarations}

\bmhead {Conflict of interest} The authors declare no conflicts of interest.



\bibliography{sn-bibliography}% common bib file


\end{document}